\documentclass[prl,twocolumn,preprintnumbers,amsmath,amssymb]{revtex4}

\usepackage{epsfig}
\usepackage[colorlinks=true,linkcolor=blue]{hyperref} 
\usepackage{color}
\usepackage{ulem}
\usepackage{dcolumn}
\DeclareGraphicsExtensions{.pdf}

\newcommand{\Fstr}{F}
\newcommand{\Flstr}{F_l}
\newcommand{\Fhstr}{F_h}

\begin{document}

\preprint{SMU-HEP-21-07}
\preprint{MSUHEP-21-012}

\title{General heavy-flavor mass scheme for charged-current DIS at NNLO and beyond}

\author{Jun~Gao$^{1,2,3}$,
        T.~J.~Hobbs$^{4,5,6,7}$,
        P.~M.~Nadolsky$^6$,
	    ChuanLe Sun$^{1,2}$,
        C.-P.~Yuan$^8$}

\affiliation{
    $^1$ INPAC, Shanghai Key Laboratory for Particle Physics and Cosmology \\ \& School of Physics and Astronomy, Shanghai Jiao Tong University, Shanghai 200240, China \\
    $^2$Key Laboratory for Particle Astrophysics and Cosmology (MOE), Shanghai 200240, China \\
    $^3$Center for High Energy Physics, Peking University, Beijing 100871, China \\
    $^4$Fermi National Accelerator Laboratory, Batavia, IL 60510, USA \\
    $^5$Department of Physics, Illinois Institute of Technology,
 Chicago, IL 60616, USA \\
	$^6$\mbox{Department of Physics, Southern Methodist University,
 Dallas, TX 75275-0175, USA} \\
    $^7$\mbox{Jefferson Lab, EIC Center, Newport News, VA 23606, USA} \\
    $^8$Department of Physics and Astronomy, Michigan State University, East Lansing, MI 48824, USA}

\date{\today}

\begin{abstract}
 Incompleteness in current knowledge of neutrino interactions with nuclear matter imposes a primary limitation in searches for leptonic CP violation carried out at long-baseline neutrino experiments. In this paper, we present a new computation that elevates the theoretical accuracy to next-to-next-to-leading order (NNLO) in QCD for charged-current deeply-inelastic scattering (DIS) processes relevant for ongoing and future neutrino programs. Mass-dependent quark contributions are consistently included across a wide range of momentum transfers in the SACOT-$\chi$ general-mass scheme. When appropriate, we further include N$^3$LO corrections in the zero-mass scheme. We show theoretical predictions for several experiments with neutrinos  over a wide range of energies and at the upcoming Electron-Ion Collider. Our prediction reduces perturbative uncertainties to $\sim\!1\%$, sufficient for the high-precision objectives of future charged-current DIS measurements, and provides important theoretical inputs to experimental studies of leptonic mixing and CP violations.
\end{abstract}
\maketitle

%%%%%%%%%%%%%%%%%%%%%%%%%%%%%%%%%%%%%%%%%%%%%%%%%%%%%%%%%%%%%%%%%%%%%%%%

%
Combined charge-conjugation and parity-reversal (CP) symmetry of elementary
particles is a fundamental symmetry between matter and antimatter, and CP violation is necessary to explain the observed
imbalance in the abundances of matter and antimatter in the Universe. However, 
the observed CP violation in the quark sector is too small to account for
this imbalance by itself.
On the other hand, leptonic mixing  in charged-current
interactions remains less constrained and may provide a potential
source of CP violation.

In recent years, an ambitious international program to constrain
a possible lepton-sector CP-violating phase, $\delta_\mathrm{CP}$, 
has been pursued in muon- to electron-(anti)neutrino oscillation searches
at a variety of facilities, 
including the Tokai-to-Kamioka (T2K)~\cite{1807.07891}, NOvA~\cite{1906.04907}
and DUNE~\cite{1512.06148} experiments.
To determine whether leptonic CP violation is large enough to account for the matter-antimatter asymmetry, these experimental tests require tight theoretical control over 
charged-current production rates, which in turn
entails a global effort to advance the associated nuclear and hadronic models, as well as
perturbative QCD computations~\cite{1706.03621}.
This effort, as well as experimental programs to constrain $\delta_\mathrm{CP}$ in
the lepton sector, run parallel to an experimental-theoretical campaign
to explore neutrino-nucleus interactions to higher precision
in short-baseline neutrino experiments or at the near detectors of long-baseline searches for CP violation.
In this setting, theoretical understanding of charged-current (CC) deeply-inelastic
scattering (DIS) is essential not
only because this process dominates at
high energies but also because determination of the
neutrino flux, including its energy dependence and overall normalization, relies strongly
upon modeling of CC DIS~\cite{Seligman:1997fe,Fine:2020knh}. 

Charged-current deeply-inelastic scattering has the potential to
unlock unique combinations of quark-flavor currents inside QCD matter and
is therefore a useful complement to neutral-current (NC) DIS as a probe of hadronic
and nuclear structure. 
There have been numerous CC DIS measurements from fixed-target experiments
(see Ref.~\cite{Zyla:2020zbs} for an overview) as well as from HERA~\cite{Abramowicz:2015mha}.
In addition, as a dominant contribution to the total inclusive CC cross section for (anti)neutrino scattering off nuclei
at $E_\nu\! \sim\! [\mathrm{few\,\, GeV}]$ and beyond,
CC DIS plays an essential role in various neutrino experiments,
including the long-baseline programs noted above as well as
the IceCube neutrino telescope~\cite{Ahrens:2003ix} and FASER$\nu$~\cite{2001.03073}
at the LHC.
An enhanced theoretical understanding of the neutrino DIS cross section will therefore
advance the precision objectives of several neutrino experiments operating over a
wide energy spectrum.
Such theoretical advancements will be also relevant for the future
Electron-Ion Collider (EIC)~\cite{Accardi:2012qut,AbdulKhalek:2021gbh,Arratia:2020azl},
which, like HERA, will exploit CC DIS to explore the flavor dependence of hadrons' three-dimensional
structure.

In perturbative calculations of QCD, 
control over heavy-quark (HQ) contributions is vital to achieving high-precision
in theoretical calculations of DIS cross sections~\cite{Witten:1975bh,Barnett:1987jw,Olness:1987ep,Hou:2019efy,Xie:2019eoe}.
At lower energies, inclusion of threshold effects
from heavy quarks is mandatory, while all-order resummation of logarithms of heavy-quark masses
is needed at energies much larger than the masses.
A uniform description of both effects is thus desirable, given the wide span
of neutrino energies in above experiments.
In this article, we advance theoretical accuracy in electroweak physics by employing the Simplified-ACOT-$\chi$ (SACOT-$\chi$) general-mass (GM) 
scheme \cite{Aivazis:1993kh,Aivazis:1993pi,Collins:1998rz,Kramer:2000hn,Tung:2001mv}
to present the first calculation of inclusive CC DIS at 
next-to-next-to-leading order (N$^2$LO)
in QCD with full mass dependence. 
In the counting prescription that we adopt, N$^2$LO and N$^3$LO approximations include up to 2 and 3 QCD loops in CC DIS Wilson coeffcients, respectively.
GM variable-flavor number schemes \cite{Aivazis:1993kh,Aivazis:1993pi,Collins:1998rz,Kniehl:1995em,Cacciari:1998it, Buza:1996wv,Chuvakin:1999nx,Thorne:1997ga,Kramer:2000hn,Tung:2001mv} interpolate
between the two extremes
noted above, in which a GM calculation matches onto a 
{\it fixed-flavor number} (FFN) 
scheme
at momentum transfer $Q\!\sim\! M_Q$ while converging to a zero-mass (ZM) scheme
at $Q\! \gg\! M_Q$, thereby fully specifying the relevant cross section
over a broad range of $Q$~\cite{Alekhin:2009ni,Forte:2010ta,Cacciari:2012ny,Kniehl:2011bk,Kniehl:2012ti,Helenius:2018uul}.
Derived from the all-orders proof of QCD factorization for DIS with massive
quarks \cite{Collins:1998rz}, the SACOT-$\chi$ scheme offers crucial advantages:
simpler implementation of mass dependence, stable perturbative convergence,
and control of partonic threshold effects. 
In particular, while the dominant mass-depedent terms at large virtualities are known to N$^2$LO \cite{Buza:1997mg,Blumlein:2014fqa} and even N$^3$LO \cite{Behring:2014eya}, 
our calculation includes the N$^2$LO mass-dependent terms exactly and hence also predicts the threshold behavior of CC DIS cross sections.
Below, we outline the SACOT-$\chi$ theoretical framework for CC DIS and
apply it to several phenomenological studies.

\begin{figure}[ht!]
\includegraphics[width=0.2\textwidth]{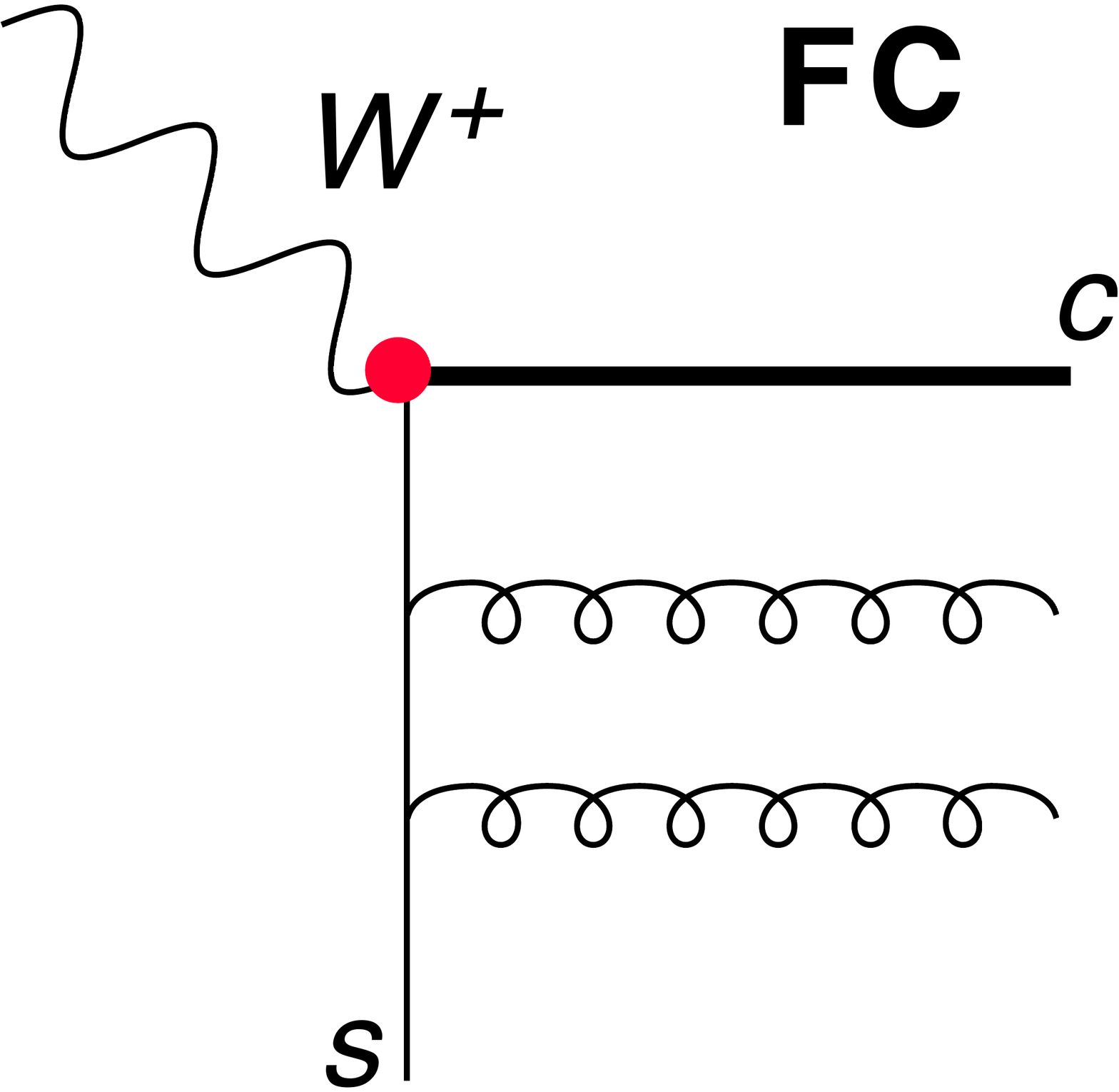} \ \ \ \ \
\includegraphics[width=0.2\textwidth]{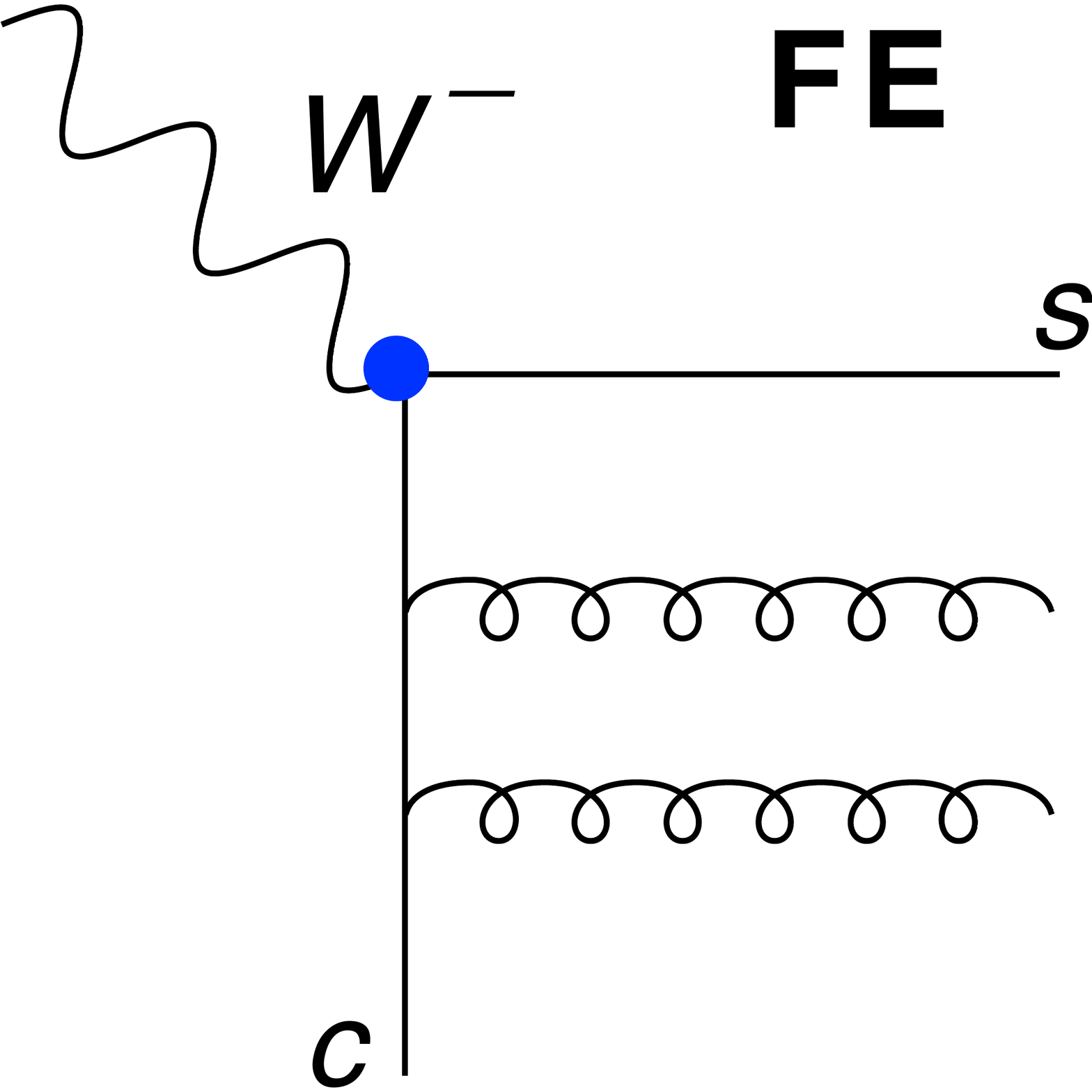}
\caption{
Representative CC DIS diagrams at N$^2$LO for either flavor
creation (left) or excitation (right), with the latter being
effectively proportional to the HQ PDF.
}
\label{fig:FE_1}
\end{figure}
%

%

%
%

%%%%%%%%%%%%%%%%%%%%%%%%%%%%%%%%%%%%%%%%%%%%%%%%%%%%%%%%%%%%%%%%%%%%%%%%

%
\section{The SACOT-$\chi$ scheme}

We proceed by extending the previous realization of the SACOT-$\chi$ scheme in Ref.~\cite{Guzzi:2011ew} for neutral-current DIS at N$^2$LO to the analogous problem in the charge-current sector,
explicitly tracing the HQ mass dependence through various radiative contributions at $\mathcal{O}(\alpha^2_s)$. 
We first demonstrate this method on the DIS structure functions, $F=F_1,\ F_2,\ F_3$, before computing DIS reduced cross sections. Up to N$^2$LO, QCD factorization allows a structure function to be written
as a convolution of parton-level coefficient functions, $C_{i,j}$, and
nonperturbative correlation functions, $\Phi$, {\it i.e.}, the parton distributions functions (PDFs), as, 
\begin{align}
	\Fstr(x,Q) &= \sum_{i}\, \sum_j \left\{ C_{i,j} \otimes \Phi_j \right\} (x,Q)\nonumber\\
	& \equiv \Flstr(x,Q)+\Fhstr(x,Q)\ ,
	\label{eq:F2conv}
\end{align}
where ``$\otimes$" denotes a convolution over the momentum fraction $z$ appearing in the expressions below, and for simplicity we do not show the electroweak (EW) couplings, 
including the Cabibbo-Kobayashi-Maskawa (CKM) matrix elements.
The equation sums over contributions from the relevant active parton flavors ($j$) in the initial state and parton flavors ($i$) produced in the final state. 
In order to implement the proper HQ mass dependence, it is necessary
to decompose the convolution in the RHS of Eq.~(\ref{eq:F2conv}) according
to the topology and flavor structure of the participating Feynman diagrams.
In this work, we 
take the maximum number of active quark flavors inside the nucleon to be $N_f=4$, together with the gluon. 

Each structure function $\Fstr(x,Q)$ is a sum of $\Flstr(x,Q)$ and $\Fhstr(x,Q)$ defined as follows:
\begin{itemize}
\item $\Flstr$ contains contributions in which only light-quark flavors ($q_l$) are directly coupled to the $W^\pm$ boson via the $W{q_l}{{\bar q}_l}$ vertex. 
\item $\Fhstr$ contains contributions involving $W{q_h}{{\bar q}_l}$ or $W{q_l}{{\bar q}_h}$ vertices. Here, $q_l$ denotes the $u$, $d$ and $s$ quarks, and  
$q_h$ the charm quark. 
\end{itemize}

Contributions to $\Flstr$ and $\Fhstr$ can be classified as representing either
{\it flavor excitation} (FE) or {\it flavor creation} (FC) depending on whether
the heavy quark appears in the initial state or only the final and virtual states.
In CC DIS, $\Flstr$ receives HQ contributions starting from N$^2$LO, while there are both FE and FC diagrams for $\Fhstr$ at LO.
Two representative Feynman diagrams for $\Fhstr$ at N$^2$LO are shown in Fig.~\ref{fig:FE_1}.
The Wilson coefficients $C_{i,j}(z)$ can be expanded in the QCD coupling $a_s\equiv \alpha_s(\mu, N_f)/(4\pi)$ as
\begin{align}
    C_{i,j}(z)= C^{(0)}_{i,j}+a_sC^{(1)}_{i,j}+a_s^2C^{(2)}_{i,j}+\mathcal{O}(a_s^3),
\end{align}
with the LO coefficients given by
\begin{align}
C^{(0)}_{l,l}&=\delta(1-z),\,\,C^{(0)}_{h,h}=\delta(1-\chi),\nonumber \\
C^{(0)}_{h,l}&=\delta(1-\chi),\,\, \chi\equiv (1+m_c^2/Q^2)z,
\end{align}
where $C^{(0)}_{h,l}$ and $C^{(0)}_{h,h}$ correspond to FC and FE contributions, respectively.
At NLO, there are gluon contributions to $\Flstr$ and $\Fhstr$,
\begin{align}
C^{(1)}_{l,l}&=c^{(1)}_{l,l}(z),\,\,C^{(1)}_{l,g}=c^{(1)}_{l,g}(z),\,\,C^{(1)}_{h,h}=c^{(1)}_{l,l}(\chi),\nonumber \\
C^{(1)}_{h,l}&=H^{(1)}_l(z)-C_{h,l}^{(0)}\otimes A_{ll}^{(1)},\nonumber\\
C^{(1)}_{h,g}&=H^{(1)}_g(z)-C_{h,l}^{(0)}\otimes A_{lg}^{(1)}-C_{h,h}^{(0)}\otimes A_{hg}^{(1)}.
\end{align}
Here the lowercase coefficients $c^{(1)}_{ij}(z)$ are given by their ZM expressions \cite{hep-ph/9907472,hep-ph/0006154}.
$H^{(1)}_{l(g)}$ are the massive coefficients for CC at NLO~\cite{Gottschalk:1980rv,hep-ph/9603304,Blumlein:2011zu}, and $A_{ij}$ are the corresponding operator-matrix elements (OMEs)~\cite{hep-ph/9612398}.
Note that, in the FE contributions, $z$ has been replaced by the scaling variable $\chi$
according to the SACOT-$\chi$ convention. 

There are several complications when extending to N$^2$LO. Firstly, as mentioned,
there are now HQ contributions to $\Flstr$,
\begin{align}
    C^{(2)}_{l,g}&=c^{(2)}_{l,g}(z),\,\, C^{(2)}_{l,h}=c^{(2)}_{l,h}(\chi),\nonumber \\
    C^{(2)}_{l,l}&=c^{(2)}_{l,l}(z)+\widetilde{C}^{(NS,2)}_{l,l}(z),
\end{align}
where the N$^2$LO ZM coefficient functions $c^{(2)}_{i,j}(z)$ are calculated in Refs.~\cite{hep-ph/9907472,hep-ph/0006154}.
$\widetilde{C}^{(NS,2)}_{l,l}$ denotes the non-singlet FC contribution after subtracting its massless counterpart, which has been included in $c^{(2)}_{l,l}(z)$ to avoid double-counting. 
The expression for $\widetilde{C}^{(NS,2)}_{l,l}$, 
with its full charm-quark mass dependence, can be found in Refs.~\cite{hep-ph/9601302,1605.05541,Behring:2015roa}. 
For $\Fhstr$, the N$^2$LO FE and FC contributions are
\begin{align}
    C^{(2)}_{h,h}&=c^{(2)}_{h,h}(\chi),\nonumber \\
    C^{(2)}_{h,l}&=H^{(2)}_{l}(z)- \Delta C^{(2)}_{h,l},\nonumber \\
     C^{(2)}_{h,g}&=H^{(2)}_{g}(z)- \Delta C^{(2)}_{h,g},
\end{align}
where the two-loop massive coefficient function $H^{(2)}_{l(g)}$ is calculated numerically in Refs.~\cite{Berger:2016inr,Gao:2017kkx}. The
subtraction terms, $\Delta C^{(2)}_{h,l(g)}$, can be constructed using the
lower-order coefficient functions and the two-loop OMEs from Ref.~\cite{hep-ph/9612398}. Their
full expressions are lengthy and will be included in a forthcoming paper. 

Furthermore, the N$^3$LO ZM coefficient functions for CC DIS have recently been calculated in Refs.~\cite{hep-ph/0411112,hep-ph/0504242,hep-ph/0608307,0708.3731,1606.08907} and implemented in the numerical program HOPPET~\cite{0804.3755,1811.07906}. When $Q^2 \gg M_Q^2$, the ZM N$^3$LO Wilson coefficients serve as the precise limit for the GM N$^3$LO ones,  while at $Q^2\approx M_Q^2$ they may miss potentially important mass-dependent contributions. 
A detailed study, but for NC DIS processes, can be found in Ref.~\cite{WangBowen:2015}. With these considerations, we also compute an {\it approximate} N$^3$LO prediction, called GM N$^3$LO$^\prime$, by adding the ${\cal O}(\alpha_s^3)$ ZM contributions without $\chi$ rescaling to the GM N$^2$LO Wilson coefficients and using N$^2$LO PDFs.

%%%%%%%%%%%%%%%%%%%%%%%%%%%%%%%%%%%%%%%%%%%%%%%%%%%%%%%%%%%%%%%%%%%%%%%%
%%%%%%%%%%%%%%%%%%%%%%%%%%%%%%%%%%%%%%%%%%%%%%%%%%%%%%%%%%%%%%%%%%%%%%%%

%
\section{Phenomenology}
We now summarize several phenomenological studies with our predictions incorporating full charm-quark mass effects.
We use CT14 NNLO PDFs \cite{Dulat:2015mca} with up to 3 active quark flavors for FFN predictions, and up to 4 active flavors with ZM and GM   
predictions. The charm-quark pole mass is taken to be 1.3 GeV, and the
CKM matrix elements are chosen according to Ref.~\cite{Zyla:2020zbs}, 
in which the third generation is assumed to be diagonal.
For the EW parameters, we use the $G_{F}$
scheme~\cite{Denner:1990ns}. 
We set the renormalization and factorization scales to the momentum
transfer, $\mu_R\! =\! \mu_F\! =\! Q$, unless otherwise specified.
After briefly considering our GM scheme for a generic reduced
cross section, we highlight specific applications to neutrino-nucleus
DIS and envisioned high-$Q^2$ measurements at the future EIC. 
%

%%%%%%%%%%%%%%%%%%%%%%%%%%%%%%%%%%%%%%%%%%%%%%%%%%%%%%%%%%%%%%%%%%%%%%%%

\subsection{A generic lepton-proton reduced cross section}
\begin{figure}[ht]
\centering
\includegraphics[width=0.5\textwidth]{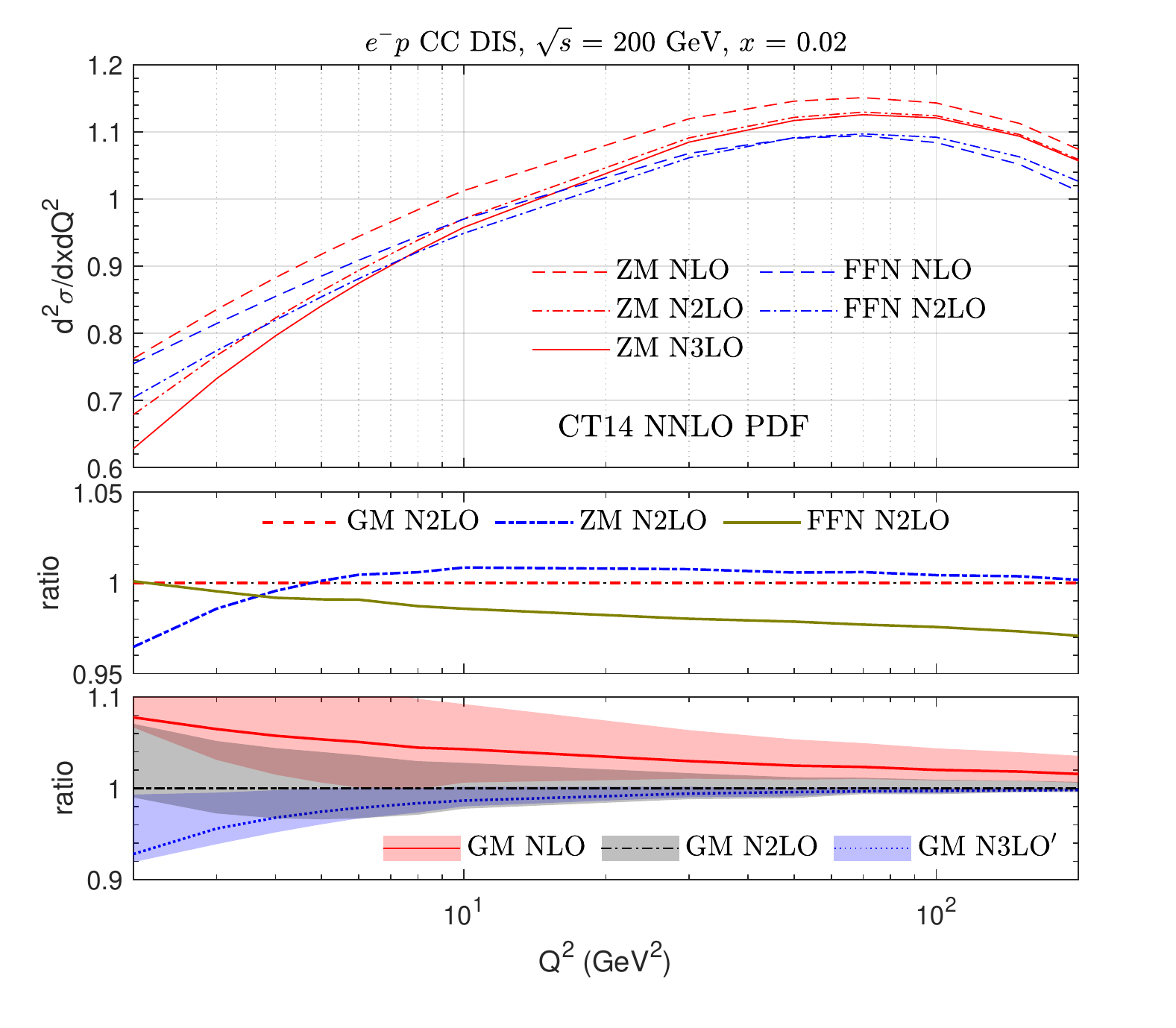}
\caption{
$Q^2$ dependence of differential reduced cross sections in $e^-p$ CC DIS at $\sqrt{s}=200$ GeV for $x=0.02$. Colored bands represent scale variations described in the main text.
\label{fig:toy1}}
\end{figure}

Figure~\ref{fig:toy1} compares predictions within the ZM, FFN and GM schemes for a reduced differential cross section $d^2\sigma/(dxdQ^2)$ at a typical Bjorken-$x$ value of 0.02.
The upper panel shows predictions for $e^-p \to \nu_e X$ from NLO up to the highest available orders in the FFN and ZM schemes for $2\, \mathrm{GeV}^2\! \le\! Q^2\! \le\! 200\, \mathrm{GeV}^2$.
At NLO, differences among the FFN and ZM predictions can reach 6\% for high-$Q^2$ values. At N$^2$LO, $\lesssim\! 3\%$ differences persist at both the lowest and highest $Q^2$. 
The middle panel compares N$^2$LO predictions for three schemes by showing ratios to the prediction of the GM scheme.
Evidently, the GM prediction interpolates nicely between the ZM and FFN predictions over a wide $Q^2$ interval.
The lower panel shows ratios of the GM predictions with scale variations at various $\alpha_s$ orders. 
The scale variations are calculated by varying $\mu_R$ and $\mu_F$ simultaneously by a factor of two, while keeping them above the charm-quark mass. 
The N$^2$LO prediction in the denominators assumes the nominal $\mu_{R,F}=Q$ scales.
The GM results converge well, and the scale dependence decreases prominently with the $\alpha_s$ order. The scale dependence is truly small at high $Q^2$ -- about 1\% for GM N$^2$LO and just a few per mille for GM N$^3$LO$^\prime$.
At $Q^2 < 10\mbox{ GeV}^2$, the GM N$^2$LO scale dependence of up to 5\% remains substantial, in fact covering the differences between the three schemes at this order. 
The partial GM N$^3$LO$'$ prediction does not include the ${\cal O}(\alpha_s^3)$ mass terms essential near the charm mass threshold, and in fact it need not be convergent at $Q^2\sim M^2_Q$, yet it yields a smaller scale variation even at low Q.

In these calculations, we neglected contributions from bottom quarks, although their inclusion in our SACOT-$\chi$ formalism is straightforward. At high $Q^2$, $b$-quark pairs can be produced in CC DIS via N$^2$LO corrections, and there are virtual $b$-quark loops in gluon self-energy subgraphs.
For a $Q^2$ value of 200 GeV$^2$, and with the same setup
as in Fig.~\ref{fig:toy1}, we found the bottom quark contributions to be small, about one per mille of the total cross section for a wide range of $x$ values. 
%

%%%%%%%%%%%%%%%%%%%%%%%%%%%%%%%%%%%%%%%%%%%%%%%%%%%%%%%%%%%%%%%%%%%%%%%%

%
\subsection{Neutrino DIS}

\begin{figure}[ht]
\centering
\includegraphics[width=0.5\textwidth]{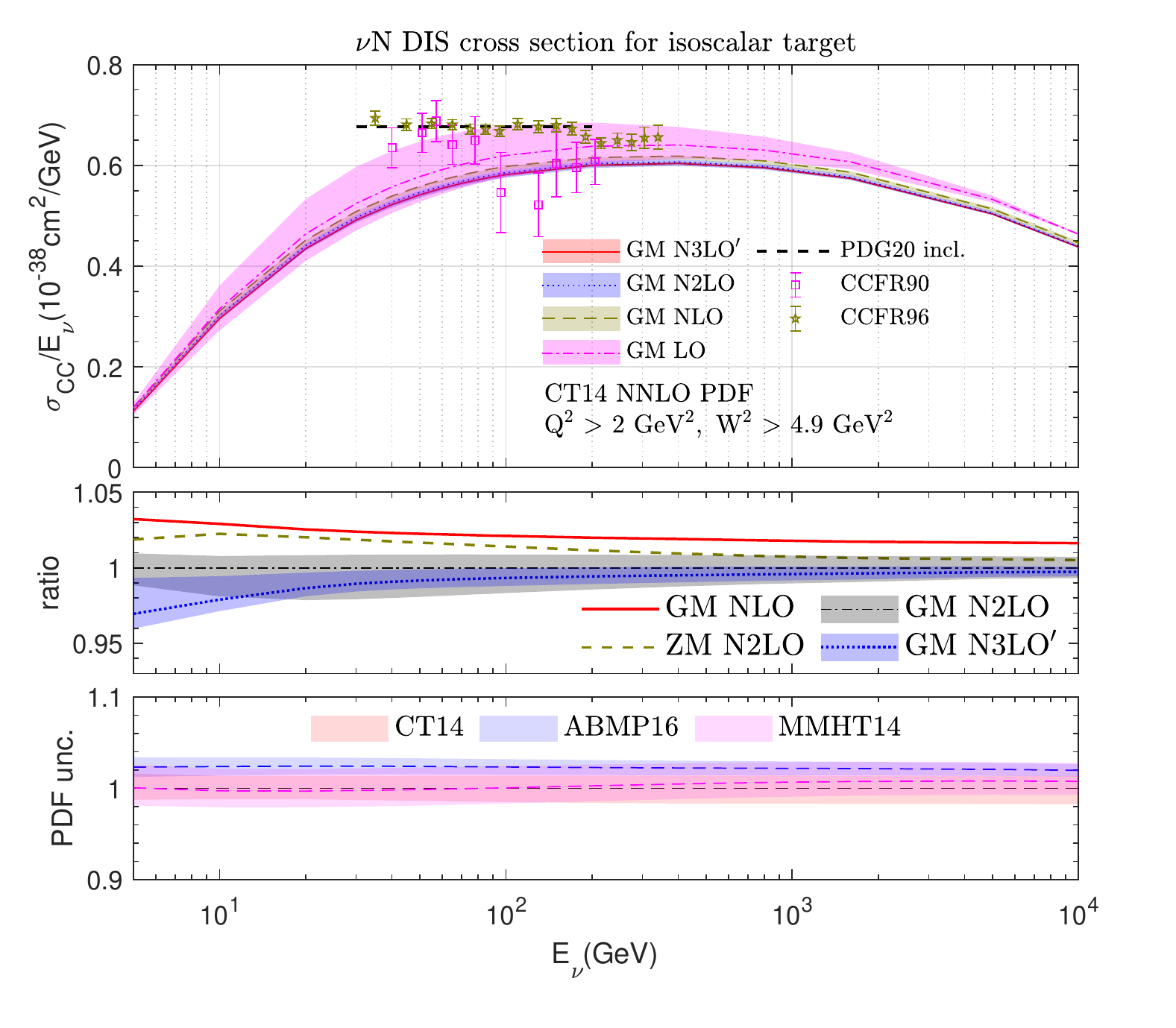}
\caption{
Curved lines: the predicted CC DIS cross section in the SACOT-$\chi$ scheme at various orders versus the neutrino energy, $E_\nu$.
Error bars and dashed horizontal line: CCFR measurements and the world average of the neutrino-nucleus total cross section. 
 Colored bands in the upper/middle (lower) panel represent the scale variations (PDF uncertainty).
\label{fig:nu1}}
\end{figure}

Next, we turn to the inclusive CC DIS cross section for neutrino scattering off an isoscalar nuclear target as a function of the neutrino energy, $E_\nu$. Figure~\ref{fig:nu1} compares experimental measurements of neutrino-nucleus {\it total} inclusive cross sections divided by $E_\nu$ to
our predictions for CC DIS cross sections for $E_\nu$ ranging from 5 to 10$^4$ GeV. We require $Q^2\!>\!2\,{\rm GeV^2}$ and $W^2\!>\!4.9\,{\rm GeV^2}$.
Many completed and upcoming fixed-target experiments have  $E_\nu<400$ GeV. At very low $E_\nu$, the measured total cross section receives sizable quasi-elastic scattering and resonant production contributions \cite{1706.03621} on top of the DIS component that we compute.
We stress that, even at lower $E_\nu$, as in long-baseline experiments like DUNE~\cite{1512.06148}, the CC DIS contribution remains important, accounting for more than $\!40\%$ of the total event rate for $E_{\nu}\sim 10$ GeV. As such, a few-percent correction to the DIS subprocess can be consequential to the ultimate precision of flavor-oscillation searches. DUNE, for instance, aims for percent-level precision in its neutrino oscillation search program.
At high neutrino energies above 100 GeV, CC DIS dominates.  The higher values of $E_\nu$ considered here can be accessed at FASER$\nu$~\cite{2001.03073} and IceCube~\cite{1711.11043}.

In Fig.~\ref{fig:nu1}, the world-average value of $\sigma_{CC}/E_{\nu}$ as reported in PDG20,  
$0.677\pm 0.014$~\cite{Zyla:2020zbs},
was originally documented in Ref.~\cite{Seligman:1997fe} by combining the CCFR90~\cite{Auchincloss:1990tu},
CCFRR~\cite{Blair:1983su}, and CDHSW~\cite{Berge:1987zw} measurements with $E_{\nu}$ between 30 to 200 GeV. This is displayed as the black dashed line.
The CCFR90~\cite{Auchincloss:1990tu} measurements extract the total cross sections with an
independent determination of the neutrino flux.  
On the other hand,  
CCFR96~\cite{Seligman:1997fe}, like 
many other neutrino-scattering experiments, only 
measured relative cross sections to cancel the neutrino flux uncertainty. 
The reported absolute cross sections as a function of $E_{\nu}$,
$\sigma_{CC}/E_{\nu}$, were obtained by matching onto the above-mentioned world-average value.
We note that in recent accelerator-based neutrino experiments, e.g., NuTeV~\cite{hep-ph/0509010}, 
NOMAD~\cite{0711.1183}, MINOS~\cite{0910.2201}, MINERvA~\cite{1610.04746}, the absolute neutrino
fluxes are all normalized using the same world-average value.

Our theory predictions include NLO EW corrections, 
as originally calculated in Ref.~\cite{hep-ph/0407203}, and nucleon-level target mass corrections
following the prescription of Ref.~\cite{Georgi:1976ve}.
For $E_\nu\! =\! 200$ GeV, these corrections increase the DIS cross section by about 2\% and 1\%, respectively. 
Furthermore, we check nuclear-to-isoscalar corrections using the nCTEQ15 PDFs~\cite{1509.00792}, finding
these only decrease cross sections by $<\!0.5\%$, assuming $A\!=\!56$ for an iron nucleus.
The upper panel of Fig.~\ref{fig:nu1} shows 
the GM theory predictions at LO, NLO, N$^2$LO and N$^3$LO$^\prime$, as well as the ZM prediction at N$^3$LO.
QCD corrections reduce the LO cross sections by about 6\% for
most neutrino energies.
The scale dependence indicated by the colored band is strongly reduced upon including higher-order corrections. 
The middle panel of Fig.~\ref{fig:nu1} further compares theoretical predictions obtained at various QCD orders by examining ratios to the GM N$^2$LO cross section. 
The scale variation for GM N$^2$LO and especially N$^3$LO$^\prime$ is 
negligible at $E_\nu > 100$ GeV and is 1-3\% otherwise.
One important feature is that higher-order QCD corrections reduce the DIS cross section to somewhat increase the apparent difference between the precise CCFR96 data and theory predictions, which 
may be attributed to the low-$Q^2$ contributions that are not included in this theory calculation. 
The agreement with the CCFR90 data is better, especially for $E_{\nu} $ above $\approx$100 GeV.
The ambiguity due to the absent mass terms grows up to a few percent in the ZM and GM N$^3$LO$^\prime$
predictions for the lowest $E_\nu$. This ambiguity is reduced in GM N$^2$LO.  
These differences can be contrasted with the PDF uncertainties in the range 1$\sim$2\% in the lower panel of Fig.~\ref{fig:nu1}.
We also compare N$^2$LO predictions using a few other PDF sets, MMHT2014~\cite{1412.3989}
and ABMP16~\cite{1701.05838}, calculated with the ZM scheme of 4 flavors
and with PDF uncertainties at 68\% C.L.
They agree with CT14 predictions within the PDF uncertainty, as shown in
the lower panel for MMHT2014 and ABMP16.

%

%%%%%%%%%%%%%%%%%%%%%%%%%%%%%%%%%%%%%%%%%%%%%%%%%%%%%%%%%%%%%%%%%%%%%%%%

%
\subsection{HERA/EIC kinematics}
\begin{figure}[ht]
\centering
\includegraphics[width=0.5\textwidth]{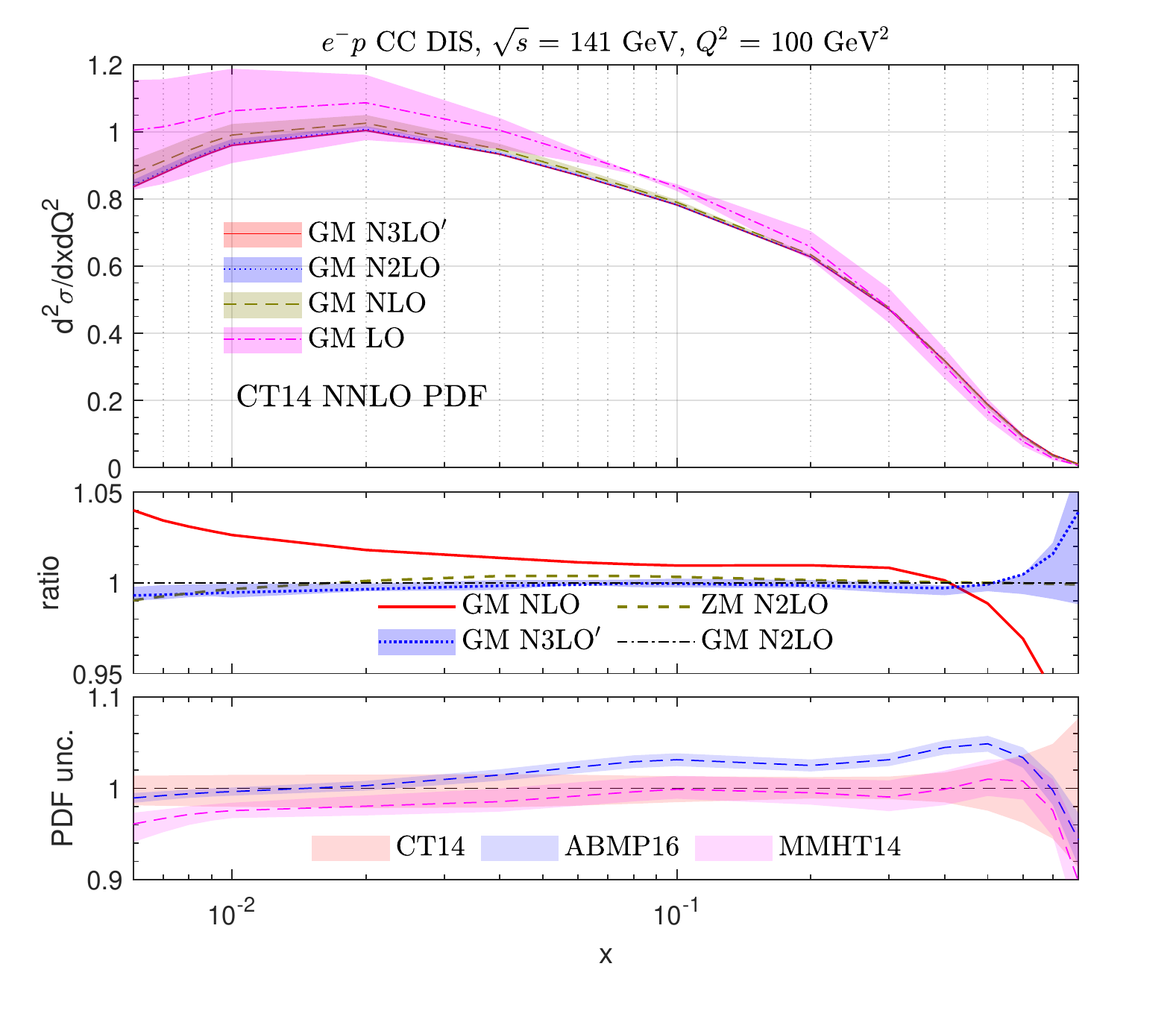}
\caption{Bjorken-$x$ dependence of differential reduced cross sections in $e^-p$ CC DIS at $\sqrt{s}=141$ GeV for $Q^2=100\,{\rm GeV^2}$.
Colored bands in the upper/middle (lower) panel represent the scale variations (PDF uncertainty).
\label{fig:eic1}}
\end{figure}

Inclusive CC DIS can be measured precisely at the future EIC~\cite{Accardi:2012qut,AbdulKhalek:2021gbh,Arratia:2020azl}.
At lepton-hadron colliders like HERA and the EIC, the typical $Q^2$ in CC DIS is above 100 ${\rm GeV^2}$ due to difficulties of reconstructing the full
hadronic energy~\cite{1708.01527,Arratia:2020azl}.
Figure~\ref{fig:eic1} shows reduced cross sections and ratios vs.~$x$ at $Q^2=100\,{\rm GeV^2}$
for $e^-p$ collisions with a center-of-mass energy of 141 GeV. The comparison of GM predictions at various $\alpha_s$ orders, including their scale variations, again demonstrates good perturbative convergence.
At such $Q^2$, GM N$^3$LO$^\prime$ is an excellent prediction, as the charm-mass terms are negligible. The GM N$^3$LO$^\prime$ scale dependence is within 0.5-1\%, except at very large $x$. By comparing the GM and ZM predictions, we find that the full charm-quark mass effects can still lead to a correction of $\approx$1\%, depending on the $x$ values.
Such high theoretical accuracy represents another step toward higher-precision tests of QCD in CC DIS at the EIC.
Meanwhile, the PDF uncertainties based on CT14 in the lower panel are generally about 2\%.
We note that the ABMP16 predictions can differ from CT14 by almost 4\% in the large-$x$ region.

%

%%%%%%%%%%%%%%%%%%%%%%%%%%%%%%%%%%%%%%%%%%%%%%%%%%%%%%%%%%%%%%%%%%%%%%%%
%%%%%%%%%%%%%%%%%%%%%%%%%%%%%%%%%%%%%%%%%%%%%%%%%%%%%%%%%%%%%%%%%%%%%%%%

In conclusion, we have presented a general-mass calculation for inclusive CC DIS at N$^2$LO in QCD with full threshold dependence on the charm-quark mass. 
The GM N$^2$LO predictions are consistent across a wide range of momentum transfers and have greatly reduced perturbative uncertainties.
When appropriate, we augment the GM N$^2$LO calculation by including the ${\mathcal O}(\alpha_s^3)$ radiative contributions available in the zero-mass scheme. Our examination of phenomenological implications for several experimental programs, including neutrino experiments at various energies and the EIC, shows that perturbative uncertainties can be controlled at the level of a few percent and sometimes less.
In particular, our precision calculations for CC DIS --- one of the main detection processes for high-energy neutrino experiments --- provides essential theoretical input to studies of leptonic mixing and CP violation.

%%%%%%%%%%%%%%%%%%%%%%%%%%%%%%%%%%%%%%%%%%%%%%%%%%%%%%%%%%%%%%%%%%%%%%%%
\quad \\
This work is partially supported by the U.S.~Department of Energy under Grant No.~DE-SC0010129 (at SMU), and by the U.S.~National Science Foundation
under Grant No.~PHY-2013791 (at MSU). T.~J.~Hobbs acknowledges support from a JLab EIC Center Fellowship, and 
the Fermi National Accelerator Laboratory, managed and operated by Fermi Research Alliance, LLC under Contract No.~DE-AC02-07CH11359 with the U.S.~Department of Energy.
The work of J.G.~was supported by the National Natural Science Foundation of China under Grants No.~11875189 and No.~11835005. C.-P.~Yuan is also grateful for the support from the Wu-Ki Tung endowed chair in particle physics.

%%%%%%%%%%%%%%%%%%%%%%%%%%%%%%%%%%%%%%%%%%%%%%%%%%%%%%%%%%%%%%%%%%%%%%%%%
\bibliographystyle{apsrev}
\bibliography{vfncc}

\end{document}